\documentclass[preprint,pra,showpacs]{revtex4}

\usepackage{amsmath,graphicx,epsfig,bm}
 \begin{document}
 
\title{Bose-Einstein condensate in a rapidly rotating non-symmetric trap}
\author { Alexander L.\ Fetter} 
\affiliation {Departments of Physics and Applied Physics, Stanford University,
   Stanford, CA 94305-4045, USA}

\date{\today}

\begin {abstract}

 A rapidly rotating  Bose-Einstein condensate in a {symmetric} two-dimensional harmonic trap can be  described with the lowest Landau-level  set of single-particle states.  The  condensate wave function $\psi(x,y)$ is a Gaussian $\propto \exp(-r^2/2)$, multiplied by an analytic function $f(z) $ of the  complex  variable $z= x+ i y$.   The criterion for a quantum phase transition to a non-superfluid correlated many-body state is usually expressed in terms of the ratio of the number of particles to the number of vortices.  Here, a similar  description applies to  a rapidly rotating { non-symmetric} two-dimensional trap with arbitrary quadratic anisotropy ($\omega_x^2 < \omega_y^2$).  The corresponding condensate wave function $\psi(x,y)$ is a complex anisotropic Gaussian with a phase proportional to $ xy$, multiplied by an analytic function $ f(z)$, where $z = x + i \beta_- y$ is a stretched complex variable  and $ 0\le \beta_- \le 1$ is  a real   parameter that depends on the trap anisotropy and the rotation frequency.    Both in the mean-field Thomas-Fermi approximation and in the mean-field  lowest Landau level approximation with many visible vortices, an anisotropic parabolic density profile  minimizes the energy.  An elongated condensate  grows  along the soft trap direction yet ultimately shrinks along the tight trap direction.  The criterion for the  quantum phase transition to a correlated state is generalized  (1)  in terms of $N/L_z$, which suggests that a non-symmetric trap should make it easier  to observe this transition or (2) in terms of a ``fragmented'' correlated state, which suggests that a non-symmetric trap should make it harder to observe this transition.  An alternative scenario involves a crossover to a quasi one-dimensional condensate without visible vortices, as suggested by Aftalion {\it et al.}, Phys.\ Rev.\  A {\bf 79}, 011603(R) (2009).
 
\end{abstract}
\maketitle

\section{Introduction}
Most theoretical studies of rapidly rotating Bose-Einstein condensates (BECs)  assume a symmetric harmonic trap with $\omega_x^2 = \omega_y^2 = \omega_\perp^2$~\cite{Linn99,Ho01,Afta05,Bloc08,Fett09}.  In this case, as the rotation frequency $\Omega$ approaches the trap frequency $\omega_\perp$, the $N$-particle condensate  expands radially to a large radius $R_0$, leading to a reduced central density.   Quantized vortices provide the angular momentum $L_z$ with an essentially  uniform vortex density $n_v\approx \pi \hbar/(M\Omega)$, where $M$ is the particle mass.

The situation can be  quite different for a non-symmetric harmonic trap with $\omega_x^2 <\omega_y^2$~\cite{Vala56,Fett74,Linn01, Fett07,Afta09}.  In this case, the condensate expands  along the weak trap direction (here $x$),  which  enhances the  moment of inertia and hence induces extra angular momentum.  The energy functional in the rotating frame involves $E-\Omega L_z$, so that  the optimal ground state for fixed $\Omega$ tends to maximize $L_z$.  This description emphasizes the role of the deformation of the condensate.  In some situations, the sequence of  states with increasing $\Omega$ can even reduce the number of vortices in favor of increased elongation~\cite{Linn01}.

In Sec.\ II, I briefly summarize the situation for a rapidly rotating symmetric trap, when the lowest Landau level (LLL) forms a nearly degenerate set of noninteracting single-particle states. These states serve as a variational basis for a condensate wave function $\psi_{LLL}$~\cite{Ho01,Afta05}.    The criterion for a  quantum phase transition (QPT) to a correlated non-superfluid many-body state  is usually expressed as the ``filling fraction''  $N/N_v$, where $N$ is the total number of particles and $N_v$ is the number of vortices~\cite{Coop01,Sino02,Regn03,Regn04,Coop08,Vief08}.    Section III then discusses the behavior in a non-symmetric rotating trap and relies on the corresponding but more complicated LLL single-particle states in a rotating non-symmetric trap~\cite{Fett07,Afta09}.  The rapid rotation induces an  expansion along the weak trap direction and a contraction along the strong trap direction, increasing the moment of inertia and the angular momentum $L_z$.  Section IV uses these results to estimate the critical angular velocity for the quantum phase transition (QPT)  in a non-symmetric trap.  One proposed approach relies on the ratio of the total number $N$ to the angular momentum per particle $L_z$ and suggests that it may be easier to achieve this QPT in such a non-symmetric system.  A second proposed approach relies on a comparison with a ``fragmented'' correlated many-particle state~\cite{Dali09} and suggests that it may be harder to achieve this QPT in such a non-symmetric system.  A still different  scenario~\cite{Afta09} suggests an effectively  one-dimensional condensate with a fixed minimum dimension in the strong trap direction and few or no visible vortices (see also~\cite{Sinh05, Sanc05}).  Section V considers this possible crossover in some detail. 

\section{Review of  quantum phase transition for symmetric trap}

Consider a  dilute Bose-Einstein condensate with $N$ atoms, each with mass $M$,  in   a symmetric two-dimensional harmonic trap with $V(r) = \frac{1}{2}M\omega_\perp^2 r^2 = \frac{1}{2}M\omega_\perp^2 (x^2+y^2)$.  Apply a rotation  in the positive sense with angular velocity $\Omega$.  In the limit of rapid rotation ($\Omega\to \omega_\perp$), the radial expansion renders the condensate effectively two dimensional, and it is convenient to assume that the condensate is uniform in the $z$ direction with thickness $Z$.  The original three-dimensional condensate wave function can be  rescaled as $\Psi(\bm r,z)= \sqrt{N/Z}\,\psi(\bm r)$, where $\psi$ obeys the  normalization condition $\int d^2 r\,|\psi(\bm r)|^2 = 1$.  I use dimensionless variables with $\omega_\perp$ and $\sqrt{\hbar/(M\omega_\perp)}$ as units of frequency and length. 
In this way, the one-body Hamiltonian in the rotating frame is 
\begin{equation}
\label{H0}
H_0' = \frac{p^2}{2} + \frac{ r^2}{2} - \Omega L_z,
\end{equation}
where $L_z= \hat{\bm z} \cdot \bm r\bm \times\bm p = xp_y-yp_x$ is the $z$ component of  angular momentum and $\bm p=-i\bm \nabla$.
With this choice of normalization, the Gross-Pitaevskii energy functional per particle becomes 
\begin{equation}
\label{GPE'}
E'[\psi] = \int d^2 r \,\psi^*\left( \frac{p^2}{2} + \frac{ r^2}{2} - \Omega L_z +\frac{g_{2d}N|\psi|^2}{2}\right)\psi,
\end{equation}
where $g_{2d} = 4\pi a_s /Z$  is the two-dimensional coupling constant with $a_s$ the $s$-wave scattering length (assumed positive)~\cite{note1}.

\subsection{Mean-field Thomas-Fermi regime}

The mean-field Thomas-Fermi regime occurs when the rotating condensate has a dense vortex array with well-separated vortex cores.  In this case, the density variations can be neglected, and the dimensionless momentum operator becomes the superfluid velocity $\bm v$, which closely approximates solid-body rotation $\bm v_{\rm sb} = \bm \Omega \bm \times \bm r$.  Minimization of the resulting approximate Eq.~(\ref{GPE'}) with respect to the density yields the Thomas-Fermi (TF) profile
\begin{equation}\label{nTF}
|\psi(r)|^2 = n(0) \left(1-\frac{r^2}{R_{TF}^2}\right),
\end{equation}
where $n(0) = \mu/(Ng_{2d}) =2/(\pi R_{TF}^2)$ is the central density, $\mu$ is the chemical potential that enforces the normalization condition, and  the  condensate radius is given by
\begin{equation}\label{RTF}
R_{TF}^2 = 4\sqrt{\frac{Na_s}{Z\left(1-\Omega^2\right)}}.
\end{equation}
As expected physically, the condensate expands with increasing rotation, and  the central density correspondingly decreases.  The  criterion for the validity of the mean-field TF approximation is $\mu = n(0) Ng_{2d} \gtrsim 1$, which here implies
 \begin{equation}\label{MFTF}
2\sqrt{\frac{Na_s}{Z}\left(1-\Omega^2\right)}\gtrsim 1.
\end{equation}

\subsection{Mean-field lowest Landau level regime}
In the opposite limit $n(0)Ng_{2d}\lesssim 1$, the system enters the mean-field lowest Landau level (LLL) regime, when the vortex cores start to overlap.  The density variation now  becomes important, and Ho~\cite{Ho01} proposed a different approach based on the exact solution of Eq.~(\ref{H0}).
 Define dimensionless bosonic operators $a_x=(x+ip_x)/\sqrt 2$ and similarly for $a_y$~\cite{Cohe77,Fett09}.  
The one-body Hamiltonian in Eq.~(\ref{H0}) is readily diagonalized  with the following operators  \begin{equation}
\label{operators}
a_\pm = \frac{1}{\sqrt2}\left(a_x\mp ia_y\right)\quad\hbox{and}\quad a_\pm^\dagger = \frac{1}{\sqrt2}\left(a_x^\dagger\pm ia_y^\dagger\right)
\end{equation}
that destroy and create one quantum with frequency $\omega_\pm = 1\mp\Omega$ and angular momentum $\pm 1$.
The Hamiltonian becomes
\begin{equation}
\label{H0'}
H'_0 = 1 + \omega_+a_+^\dagger a_+  + \omega_-a_-^\dagger a_-,
\end{equation}
 and the angular momentum has a corresponding simple intuitive form $L_z =a_+^\dagger a_+  -a_-^\dagger a_-$.

In this helicity basis, the operator identity  $a_\pm\varphi_{00} = 0$ gives  the ground state $\varphi_{00} \propto  e^{-r^2/2}$.  For rapid rotation  $\Omega\to 1$, the  positive-helicity   normal-mode frequency $\omega_+= 1 -\Omega$ is small, whereas the negative-helicity normal-mode frequency $\omega_-= 1+\Omega \approx 2$ is of order unity.  The lowest Landau level is the set of low-lying states with with $\langle a_+^\dagger a_+\rangle = m$ and $\langle a_-^\dagger a_-\rangle= 0$.  The corresponding (normalized) low-lying excited states are 
\begin{equation}
\label{LLL}
\psi_m(x,y)  = \frac{z^m}{\sqrt{\pi m!}}\,e^{-r^2/2},
\end{equation}
where $z=  x+iy$ is a complex variable.

For this symmetric trap, the projector onto the lowest Landau level (LLL) follows directly from these normalized wave functions 
\begin{equation}
\label{Pi0}
\Pi_{0}(x,y;x',y') = \sum_{m=0}^\infty \psi_m(x,y)\,\psi_m^*(x',y'), 
\end{equation}
and a straightforward calculation yields
\begin{equation}
\label{LLLPi0}
\Pi_{0}(x,y;x',y') =  \frac{1}{\pi} \exp\left(zz'^*\right )\,\exp\left(-\frac{1}{2}|z|^2-\frac{1}{2}|z'|^2\right).
\end{equation}
The projection of any reasonable variational trial state $\psi_v(x,y)$  onto the LLL basis 
\begin{equation}
\label{psi0}
\psi_{v0}(x,y) \equiv \int d^2r'\,\Pi_0(x,y;x'y') \,\psi_v(x',y')
\end{equation}
has the form $\psi_{v0}(x,y) = f(z) \exp(-|z|^2/2)$,  where $f(z)$ is an analytic function of $z = x + i y$, and the remaining factor is effectively the ground-state wave function.  Unless $\psi_v$ already is in the LLL manifold, $\psi_{v0}$  typically has a reduced normalization.  

Assume a normalized linear combination of LLL solutions with $\psi_{LLL} =\sum_m c_m \psi_m$.  The general properties of these LLL eigenstates yield the important and simple result~\cite{Ho01}
\begin{equation}
\label{L0}
\langle L_z\rangle = \langle r^2\rangle -1,
\end{equation}
where $\langle \cdots \rangle $ means an average with $\psi_{LLL}$.
For these LLL states, the mean-field GP energy functional (\ref{GPE'}) has a simple form~\cite{Afta05,Fett09,note1a}
\begin{equation}
\label{E'LLL}
E'[\psi_{LLL}] = \Omega + \int d^2 r\left[\left(1-\Omega\right) r^2\,|\psi_{LLL}|^2 + \frac{1}{2}g_{2d} N\, |\psi_{LLL}|^4\right].
\end{equation}
For numerical estimates, I use $a_s/Z \approx 5\times 10^{-3}$ and $N=5\times 10^3$, so that $g_{2d}N = 4\pi N a_s/Z \approx 300$, in qualitative agreement with one relevant
experimental   coupling constant $4\pi a_s  \tilde{N} \sim 130$~\cite{Rose02}, where $\tilde N$ is the number of particles per unit axial length.

The approximate LLL ground state follows by minimizing Eq.~(\ref{E'LLL}) for fixed particle number and gives a parabolic density profile in the radial direction 
\begin{equation}
\label{dens}
n_{LLL}(r) = |\psi_{LLL}(r)|^2 = \frac{2}{\pi R_{LLL}^2}\left( 1 - \frac{r^2}{R_{LLL}^2}\right)
\end{equation}
Here, the mean-square condensate radius 
\begin{equation}
\label{R0}
R_{LLL}^2 =\sqrt\frac{8N a_s}{Z(1-\Omega)}
\end{equation}
diverges as $1-\Omega $ becomes small~\cite{Ho01}.
I follow Aftalion, Blanc, and Lerner~\cite{Afta09} and write $1-\Omega^2 =\epsilon^2$ for this symmetric trap.  Alternatively, $1-\Omega \approx \epsilon^2 /2$, and Eq.~(\ref{R0}) can be rewritten 
\begin{equation}
\label{R0eps}
R_{LLL}^2 \approx \frac{4}{\epsilon} \sqrt\frac{Na_s}{Z},
\end{equation}
which is precisely the same as Eq.~(\ref{RTF}) for the mean-field TF regime (as seen below, the situation is slightly different for a non-symmetric rapidly rotating trap).

The validity of the LLL regime requires that the mean interaction energy $Ng_{2d} n(0)$ should be small compared to the energy gap $\omega_- \approx 2$ from the next Landau level 
\begin{equation}
\label{LLLok}
N g_{2d} n(0) =\frac{8Na_s}{ZR_{LLL}^2}\lesssim \omega_-\approx 2.
\end{equation}
A combination of Eqs.~(\ref {R0eps}) and (\ref{LLLok}) gives the criterion to achieve the LLL regime in a symmetric trap~\cite{Bloc08,Fett09}
\begin{equation}
\label{eps}
\epsilon_{LLL}^ 2\lesssim \frac{Z}{Na_s} \approx 4\times 10^{-2}
\end{equation}
for $N = 5\times 10^3$.  Equivalently, the soft frequency $\omega_+ = 1-\Omega$ has the restriction $1-\Omega \approx \frac{1}{2} \epsilon_{LLL}^2 \lesssim 2\times 10^{-2}$.  Recent JILA experiments with larger $N\approx 5\times 10^{4}$ have achieved $1-\Omega\approx 5\times 10^{-3}$, exceeding the qualitative rotation speed for the validity of the LLL regime~\cite{Schw04}.

What is the ultimate fate of such a rapidly rotating symmetric condensate?  Several authors~\cite{Coop01,Sino02,Regn03,Regn04,Coop08,Vief08} have proposed that the superfluid LLL mean-field state undergoes a quantum phase transition (QPT) to a non-superfluid correlated many-body state.  This transition occurs at a critical value of the ``filling fraction"  $N/N_v= f_c$, where  $N$ is the total number of particles, $N_v$ is the number of vortices, and $f_c$ is variously determined to be between 6 and 10 (for definiteness, I take $f_c \approx 10$). These numerical studies have used toroidal or spherical geometries with essentially no  boundaries. To apply these ideas
in the present case of a bounded circular condensate, note that each vortex occupies a dimensionless area $\pi$ for $\Omega \approx 1$, so that $N_v \approx R_{LLL}^2$.  Hence the criterion for the QPT becomes $N/R_{LLL}^2\approx f_c$, which yields the explicit critical value~\cite{Bloc08,Fett09}
\begin{equation}
\label{epsc}
\epsilon_Q^2 \approx 16\,f_c^2\frac{a_s}{NZ} \approx 1.6\times 10^{-3}.
\end{equation}


\section{Non-symmetric rotating harmonic trap}

Now consider the more general case of a  rotating non-symmetric harmonic trap~\cite{Vala56,Linn01,Okte04,Fett07,Afta09} with distinct trap frequencies $\omega_x^2<\omega_y^2$.  In the limit of rapid rotation, the condensate again becomes effectively two dimensional, and it is convenient to use the  variable $\omega_\perp^2 = (\omega_x^2 + \omega_y^2)/2$ to define both the characteristic frequency $\omega_\perp$ and the characteristic oscillator length $\sqrt{\hbar/(M\omega_\perp)}$.  With these dimensionless units, I rewrite the squared trap frequencies as $\omega_x^2 = 1-\nu^2$ and $\omega_y^2 = 1+\nu^2$, where $\nu^2 = (\omega_y^2-\omega_x^2)/2$ characterizes the splitting of the squared trap frequencies (I follow the notation of~\cite{Afta09}).

A  Bose-Einstein condensate in such a rapidly rotating non-symmetric trap tends to expand along the weak trap direction $\omega_x$, and  the condensate  elongation apparently diverges as the rotation frequency $\Omega$ tends to $\omega_x$.  Thus it is convenient to generalize the small parameter as follows:  $\epsilon^2 = \omega_x^2 -\Omega^2 = 1-\nu^2 -\Omega^2$.  For small $\epsilon$, this relation simplifies to  $\omega_x -\Omega \approx \epsilon^2/(2\omega_x)$.

\subsection{Mean-field Thomas-Fermi regime}

It is straightforward to generalize the discussion of Sec.~II.A to include a non-symmetric harmonic trap. The mean-field TF density remains parabolic in each direction
\begin{equation}\label{anisodens}
|\psi_{TF}(x,y)|^2 =n(0)\left(1-\frac{x^2}{R_{TFx}^2} - \frac{y^2}{R_{TFy}^2}\right),
\end{equation}
where $n(0) = \mu/(Ng_{2d}) =  2/(\pi R_{TFx}R_{TFy})$.  A detailed analysis yields the explicit expressions for the two squared TF condensate radii
\begin{equation}\label{RTFx}
R_{TFx}^2 = 4\sqrt\frac{Na_s}{Z}\,\frac{\left(1+\nu^2-\Omega^2\right)^{1/4}}{\left(1-\nu^2-\Omega^2\right)^{3/4}} = 4\sqrt\frac{Na_s}{Z}\,\frac{\left(\epsilon^2 + 2\nu^2\right)^{1/4}}{\epsilon^{3/2}},
\end{equation}
\begin{equation}\label{RTFy}
R_{TFy}^2 = 4\sqrt\frac{Na_s}{Z}\,\frac{\left(1-\nu^2-\Omega^2\right)^{1/4}}{\left(1+\nu^2-\Omega^2\right)^{3/4}} = 4\sqrt\frac{Na_s}{Z}\,\frac{\epsilon^{1/2}}{\left(\epsilon^2 + 2\nu^2\right)^{3/4}}.
\end{equation}
The larger TF dimension $R_{TFx}^2$ grows continuously with increasing $\Omega^2 \to 1-\nu^2$.  In contrast, the smaller TF dimension $R_{TFy}^2$ grows only for $\Omega^2  <1-2\nu^2$  (equivalently $\epsilon^2>\nu^2$).  For $\Omega^2 > 1-2\nu^2$ (equivalently,  $\epsilon^2 < \nu^2$), the smaller TF condensate dimension decreases continuously.  As discussed below, the rapidly rotating non-symmetric system eventually reaches the mean-field lowest Landau level regime, requiring a different approach.

\subsection{Classical rotating non-symmetric trap}
 
 With these dimensionless units, the one-body Hamiltonian now has the form
 \begin{equation}
\label{H1} H_0 = \frac{1}{2} \left(p_x^2 +p_y^2\right)  + \frac{1}{2}\left( \omega_x^2 x^2 + \omega_y^2 y^2\right) -\Omega\left(x p_y-yp_x\right )  
\end{equation}
where the last term couples the $x$ and $y$ motion,  The classical dynamical equations  readily yield  two normal modes with elliptical polarization.   The two frequencies are 
 \begin{equation}
 \label{modes}
\omega_\pm^2 = 1 +\Omega^2 \mp \alpha,
\end{equation}
 where $\alpha = \sqrt{\nu^4 + 4\Omega^2}$.  The plus mode  with ``small'' frequency $\omega_+$ has  positive helicity with polarization $x+i\beta_+ y$, and the minus mode with ``large'' frequency  $\omega_-$
  has negative helicity with  polarization $i\beta_-x+ y$, where \cite{Vala56,Fett07} give  detailed expressions for $\beta_\pm$. 
 
 In the present context, the most interesting  question is the behavior  for  fixed finite $\nu $ as $\Omega\to \omega_x$, when $\epsilon^2 = \omega_x^2- \Omega^2$ is small.  In this limit,   both $\omega_+$ and $\beta_+$ become small, with 
\begin{equation}
\label{plus}
\omega_+ \approx \frac{\epsilon \nu}{\sqrt{2-\nu^2}} \ \ \mbox{and}\ \  \beta_+\approx \frac{\epsilon}{\nu}\sqrt{\frac{1-\nu^2}{2-\nu^2}}.
\end{equation}
In contrast, $\omega_-$ and $\beta_-$   remain finite, with 
\begin{equation}
\label{minus}
\omega_- \approx \sqrt 2\sqrt {2-\nu^2} \ \ \mbox{and}\ \  \beta_- \approx \sqrt 2\sqrt\frac{1-\nu^2}{2-\nu^2}.
\end{equation}

\subsection{Quantum states}
The bosonic  quantum operators $A_\pm$ and $A_\pm^\dagger$ for a non-symmetric rotating trap correspond to those in Eq.~(\ref{operators}) for a symmetric trap.  They  obey the familiar bosonic commutation relations $[A_\pm,A_\pm^\dagger]=1$  with all other commutators vanishing.  Nevertheless, they are rather complicated, and the details are relegated to Appendix A [see Eqs.~(\ref{ap}) and (\ref{am})].  The normalized ground-state wave function $\varphi_{00}(x,y)$ can be defined by the relations $A_\pm \varphi_{00} = 0$; it  has the explicit form~\cite{Vala56,Fett07}
\begin{equation}
\label{gnd}
\varphi_{00}(x,y) = \frac{1}{\sqrt{\pi l_xl_y}}\,\exp\left(-\frac{x^2}{2l_x^2}-\frac{y^2}{2l_y^2} + ixy\,\frac{\alpha c-\nu^2}{2\Omega}\right),
\end{equation}
where 
\begin{equation}
\label{lxy}
l_x^2 = \frac{\Omega\left(1+\beta_+\beta_-\right)}{\alpha\beta_+},\quad l_y^2 = \frac{\Omega\left(1+\beta_+\beta_-\right)}{\alpha \beta_-},
\end{equation}
and 
\begin{equation}
\label{c}
c = \frac{1-\beta_+\beta_-}{1+\beta_+\beta_-}.
\end{equation}
Note that $c\to 0$ for a symmetric trap because in that case $\beta_+=\beta_-=1$.  In contrast, $c\to 1$ for a non-symmetric trap as $\epsilon\to 0$, since $\beta_+\propto \epsilon\to 0$.

For rapid rotation ($\epsilon\ll 1$), the frequencies $\omega_\pm$ have very different magnitude, with $\omega_+$ proportional to $ \epsilon$, and $\omega_-$ approaching a constant of order 1.  Thus the analog of the lowest Landau level in the present case follows with the  prescription 
\begin{equation}
\label{anisoLLL}
\psi_m(x,y) \equiv \varphi_{m0}(x,y) = \frac{(A_+^\dagger)^m}{\sqrt{m!}}\,\varphi_{00}(x,y),
\end{equation}
where $A_+^\dagger$ has the explicit form given in Eq.~(\ref{ap}).  A detailed analysis~\cite{Fett07} eventually yields  the LLL wave functions
\begin{equation}
\label{anisopsim}
\psi_m(x,y) = \frac{1}{\sqrt{m!}} \left(\frac{c}{2}\right)^{m/2}\,H_m\left(\sqrt{\frac{\alpha \beta_+}{\Omega c}}\,\frac{z}{1+\beta_+\beta_-}\right)\,\varphi_{00}(x,y),
\end{equation}
where $H_m$ is the usual Hermite polynomial and 
 \begin{equation}
\label{z}
z = x+i\beta_-y
\end{equation}
is a ``stretched'' complex variable. It is notable that, apart from the ground state $\varphi_{00}$, this wave function is an analytic function of the stretched complex variable $z$~\cite{Fett07,Afta09}.  Its structure is very similar to Eq.~(\ref{LLL}) for a symmetric trap, and Eq.~(\ref{anisopsim}) reduces to this elementary form when $\nu\to 0$.

The projector onto the LLL states follows directly as  in Eq.~(\ref{Pi0}).  Remarkably, this expression can be evaluated analytically using a standard formula for the sum of products of Hermite polynomials~\cite{Lebe72}
\begin{equation}
\label{hermite}
\sum_{m=0}^\infty \frac{H_m(u)\,H_m(v)}{2^m \,m!} \, c^m= W(u,v,c)\equiv \frac{1}{\sqrt{1-c^2}}\exp\left[\frac{2uvc - \left(u^2+v^2\right)c^2 }{1-c^2}\right],
\end{equation}
where 
\begin{equation}
\label{uv}
u = \sqrt{\frac{\alpha \beta_+}{\Omega c}}\,\frac{z}{1+\beta_+\beta_-}, \quad 
v = \sqrt{\frac{\alpha \beta_+}{\Omega c}}\,\frac{z'^*}{1+\beta_+\beta_-},
\end{equation}
  and $z'^* = x'-i\beta_- y'$.  Here I focus on the behavior for fixed $\nu\neq 0$ as $\epsilon$ becomes small.  Reference~\cite{Matv09} uses  a similar approach to study two limiting cases:  a circular trap  with $\nu = 0$, and a non-symmetric trap (fixed  $\nu\neq 0$) in the limit $\Omega= \omega_x$  (namely $\epsilon=0$), when the condensate becomes infinitely elongated.  
 
 When combined with the explicit ground-state wave functions from Eq.~(\ref{gnd}), a detailed calculation yields the  expression 
 \begin{equation}
\label{anisoPi0}
\Pi_0(x,y;x'y') = \frac{\alpha}{2\pi\Omega} \,\chi(x,y)\,\chi ^*(x',y')\exp\left(\frac{\alpha z z'^*}{2\Omega\beta_-}\right),
\end{equation}
where
 \begin{equation}
\label{chi}
\chi(x,y) = \exp\left(-\frac{\alpha x^2}{4\Omega\beta_-}-\frac{\alpha \beta_- y^2}{4\Omega} - i\,\frac{\nu^2 xy}{2\Omega}\right).
\end{equation}
Apart from the phase, this result is equivalent to that given in~\cite{Afta09} following a result of Bargmann~\cite{Barg61}.  In particular, for a non-symmetric rotating harmonic trap, the projection of any trial state $\psi_v(x,y)$ has the form $\psi_{v0}(x,y) =f(z)\, \chi(x,y)$~\cite{Afta09} as seen from the analogous Eq.~(\ref{psi0}).  

\subsection{Variational trial state and physical consequences}

Assume a normalized linear combination of LLL states with $\psi_{LLL} = \sum_m c_m\,\psi_m$ and $\sum_m |c_m|^2 = 1$.  The corresponding total energy functional in the rotating frame has the form~\cite{Fett07}
\begin{eqnarray}
\label{E'}
E_{LLL}[\psi_{LLL}] = \frac{1}{2}\omega_-  - \frac{1}{4} \omega_+\left( \beta_+\beta_- + \frac{1}{\beta_+\beta_-}\right)\nonumber\\[.3cm]
+\int d^2r\,\left[\frac{\alpha\omega_+}{2\Omega}\left(\beta_+ x^2+\frac{y^2}{\beta_+}\right)|\psi_{LLL}|^2 + \frac{2\pi Na_s }{Z}|\,\psi_{LLL}|^4\right].
\end{eqnarray}

Minimization of Eq.~(\ref{E'}) with respect to $|\psi_{LLL}|^2$ at fixed total number $N$ yields an anisotropic parabolic density
 \begin{equation}
\label{para}
|\psi_{LLL}(x,y) |^2 = \frac{2}{\pi R_xR_y}\left(1-\frac{x^2}{R_x^2} - \frac{y^2}{R_y^2}\right),
\end{equation}
with 
\begin{equation}
\label{RxRy}
R_x^2 = \frac{4}{\beta_+}\sqrt{\frac{Na_s\Omega}{Z\omega_+\alpha}}\quad\hbox{and}\quad R_y^2 = 4\beta_+\sqrt{\frac{Na_s\Omega}{Z\omega_+\alpha}}.
\end{equation}
In the limit of rapid rotation when $\epsilon\to 0$ and $\alpha \to 2-\nu^2$,  the squared mean-field radius along the weak trap direction 
\begin{equation}
\label{Rx}
R_x^2 \approx\frac{4\sqrt\nu}{\epsilon^{3/2}}\left(\frac{2-\nu^2}{1-\nu^2}\right)^{1/4}\sqrt\frac{Na_s}{Z}
\end{equation}
diverges strongly, proportional to  $\epsilon^{-3/2}$.  In contrast, the squared mean-field radius along the strong trap direction 
\begin{equation}
\label{Ry}
R_y^2 \approx\frac{4\sqrt\epsilon}{\nu^{3/2}}\left(\frac{1-\nu^2}{2-\nu^2}\right)^{3/4}\sqrt\frac{Na_s}{Z}
\end{equation}
shrinks slowly, proportional to $\sqrt\epsilon$.  Both of these $\epsilon$ dependences  are the same as  the leading factors in Eqs.~(\ref{RTFx}) and (\ref{RTFy}) describing the mean-field TF regime, but the overall numerical coefficients are somewhat different.

For a non-symmetric trap with finite $\nu$, the validity of the LLL regime requires that
\begin{equation}
\label{LLLOK}
Ng_{2d}n(0) =\frac{8Na_s}{ZR_xR_y} \lesssim \omega_- \approx \sqrt2 \sqrt{2-\nu^2}.
\end{equation}
A combination of the previous results yields 
\begin{equation}
\label{LLLOK1}
\epsilon_{LLL}^2 \lesssim \frac{\left(1-\nu^2\right)\left(2-\nu^2\right)}{4\nu^2}\left(\frac{Z}{ Na_s }\right)^2
\end{equation}
as the criterion for the LLL regime.  This expression scales as $N^{-2}$, in contrast to the $N^{-1}$ dependence for a symmetric trap in Eq.~(\ref{eps}).  With a  typical anisotropy parameter $\nu = 0.5$, the squared trap frequencies are $\omega_x^2= 0.75$ and $\omega_y^2 = 1.25$, leading to  $\epsilon_{LLL}^2\lesssim 2\times 10^{-3}$ for $N=5\times 10^3$ and $Z/a_s = 200$.  These numbers make the LLL regime seem less accessible for a non-symmetric trap, but reducing $N$ can increase these estimates significantly.

\section{Quantum phase transition to correlated state}

As the non-symmetric trap rotates more rapidly (with $\epsilon^2 = \omega_x^2 -\Omega^2 \ll 1$), the mean-field condensate expands  in the weakly confined direction and contracts  in the tightly confined direction [see Eqs.\ (\ref{Rx}) and (\ref{Ry})].  

\subsection{Angular-momentum criterion}

Eventually, the angular momentum $L_z$ arises in part from the increased  moment of inertia that reflects the elongation~\cite{Linn01},  along with the more familiar vortices that dominate the angular momentum for a symmetric trap.  This phenomenon occurs even for classical incompressible fluids in a rotating elliptical container~\cite{Lamb45,Fett74}.  Indeed, in one scenario, the rapidly rotating non-symmetric condensate has essentially no visible vortices~\cite{Afta09}.

This situation requires a generalization of the usual filling fraction $f$ (usually defined as the ratio $N/N_v$, which is equivalent to $N/R_0^2$ for a symmetric trap in the mean-field LLL limit).  It is easy to verify that  the dimensionless angular momentum per particle in a rapidly rotating symmetric trap is $R_0^2/3$.  Thus,  I here assume that the corresponding relevant ratio  in a non-symmetric trap is $f= N/(3L_z)$. This ratio $N/L_z$ plays a special role in analytical and numerical studies of rotating bosons with relatively small $N$~\cite{Wilk98,Wilk00}.

 In the mean-field LLL picture  with many vortices, the angular momentum per particle has a simple form~\cite{Fett07}
\begin{eqnarray}
\label{Lz}
L_z = \frac{\alpha}{2\Omega} \langle x^2+y^2\rangle + \frac{\omega_-}{2} \left(\beta_--\frac{1}{\beta_-}\right) \langle x^2-y^2\rangle -\frac{1}{2}\left(\beta_-+\frac{1}{\beta_-}\right),
\end{eqnarray}
where $\langle \cdots\rangle$ denotes an expectation value with the trial state $\psi_{LLL}$.  It  is not difficult to evaluate this quantity with the mean-field wave function in (\ref{para}), and a detailed calculation yields
\begin{equation}
\label{mflz}
L_z = \frac{\alpha-\nu^2}{12\,\Omega}R_x^2 + \frac{\alpha+\nu^2}{12\,\Omega}R_y^2 - \frac{1}{2}\left( \beta_-+\frac{1}{\beta_-}\right),
\end{equation}
 for  $\epsilon\to 0$, where $\alpha \approx 2-\nu^2$.
The first term grows proportional to $\epsilon^{-3/2}$, the last term remains constant, and the second term shrinks proportional to $\sqrt\epsilon$.  Hence the first term dominates $L_z$ as $\epsilon \to 0$ and  yields the approximate value 
\begin{equation}
\label{LLLLz}
L_z \approx \frac{2\sqrt \nu}{3\,\epsilon^{3/2}}\left(1-\nu^2\right)^{1/4}\left(2-\nu^2\right)^{1/4}\sqrt\frac{Na_s}{Z}.
\end{equation}
Compare this $\epsilon^{-3/2}$ dependence with that for a symmetric trap where $L_z\approx R_0^2/3= \frac{4}{3}\epsilon^{-1}\sqrt{Na_s/Z}$ from Eq.\ (\ref{R0eps}).

With this generalized definition of the filling fraction $f = N/(3L_z)$, it is natural to suggest that the quantum phase transition  from a coherent superfluid  state with a macroscopic condensate to a highly correlated many-body state occurs at the same critical value $f_c\approx 10$ as for a symmetric trap~\cite{Coop01,Sino02,Regn03,Regn04}.  This criterion leads to the critical rotation rate 
\begin{equation}
\label{epsc1}
\epsilon_Q^{2} \approx \left( 4\nu f_c^2\right)^{2/3} \left(1-\nu^2\right)^{1/3}\left(2-\nu^2\right)^{1/3}\left(\frac{a_s}{NZ}\right)^{2/3},
\end{equation}
which now scales as $N^{-2/3}$ instead of $N^{-1}$ for a symmetric trap [see Eq.~(\ref{epsc})].
Take $\nu = 0.5$, $Z/a_s\approx 200$, $N\approx 5\times 10^3$ (as before), and $f_c \approx 10$, which gives
\begin{equation}
\epsilon_Q^2 \approx 3.7\times 10^{-3}.
\end{equation}
Note that this ``angular-momentum'' value of $\epsilon_Q^2$ is somewhat larger than  that for the validity of the LLL approximation in Eq.\ (\ref{LLLOK1}).  Different choices for the relevant parameters (especially smaller $N$) can change these values considerably.

\subsection{Fragmentation criterion}

An alternative and quite general criterion for the quantum phase transition to a correlated many-body state relies on the following  comparison~\cite{Dali09}.   Imagine the $N$ particles distributed separately over  the $N$ lowest single particle states in the lowest Landau level (they have energies $j\omega_+$, where $j = 0,\cdots , N-1$).  This fragmented correlated state  has an energy per particle of order $N\omega_+/2$ and  is not superfluid. It also has small interaction energy because of the fragmentation.  Compare this energy  with  the interaction energy  in the lowest Landau level superfluid state  $ Ng_{2d}n(0)$.   The correlated many-body state should be favored when the fragmented energy becomes smaller than the interaction energy: $N\omega_+/2\lesssim Ng_{2d}n(0)$.

For a symmetric trap, the small frequency is $\omega_+ = 1-\Omega\approx \epsilon^2/2$.  In this case, $n(0) = 2/(\pi R_{LLL}^2)$, and a combination with Eq.~(\ref{R0eps}) yields the approximate critical rotation speed for the quantum phase transition in a symmetric trap 
\begin{equation}
\epsilon_Q^2\approx 64\,\frac{a_s}{NZ}.
\end{equation}
This value is essentially the same as that in Eq.~(\ref{epsc}) based on numerical studies of the filling fraction.

The situation in a non-symmetric trap is somewhat different, because Eq.~(\ref{plus}) shows that the small frequency in now of order $\epsilon$ (instead of $\epsilon^2$).  Furthermore, the corresponding central density is  $n(0) = 2/(\pi R_xR_y)$, so that the interaction energy becomes
\begin{equation}
Ng_{2d}n(0) \approx 2\sqrt{\nu\epsilon} \left(\frac{1-\nu^2}{2-\nu^2}\right)^{1/4} \sqrt{\frac{Na_s}{Z}}.
\end{equation}
The correlated state has a lower energy if $\epsilon \lesssim \epsilon_Q$, where 
\begin{equation}
\epsilon_Q^2\approx \frac{256}{\nu^2}\,\frac{(2-\nu^2)^3}{1-\nu^2}\,\left(\frac{a_s}{NZ}\right)^2.
\end{equation}
Note that this $N^{-2}$  dependence is similar to Eq.~(\ref{LLLOK1}) for the validity of the LLL regime in a non-symmetric trap.  
This functional dependence should be compared with that for a symmetric trap given in Eq.~(\ref{epsc}).  With the same numbers as before, this ``fragmentation'' criterion yields the unattainable value $\epsilon_Q^2\approx 7\times 10^{-9}$.   Dalibard~\cite{Dali09}  points out that this procedure  comparing the energy  of a fragmented correlated state with the interaction energy of a corresponding  condensed state applies for   other  quantum critical points beyond the present case of a rapidly rotating dilute Bose-Einstein gas~\cite{Bloc08}.

Which of these two criteria is correct?  For the symmetric trap, the mean interaction energy involves the squared radius $R_{LLL}^2$.  For a non-symmetric trap, this quantity can be generalized either as $\frac{1}{2}(R_x^2+R_y^2)$, which applies for the ``angular-momentum'' criterion, or as $R_xR_y$, which applies for the ``fragmentation criterion.''  Each gives the same approximate criterion for a condensate in a symmetric trap.  As  seen above, however, they yield  qualitatively different values for the appearance of the correlated non-superfluid state in a non-symmetric trap. Ultimately, this becomes an experimental question.

\section{Formation of an effectively one-dimensional condensate}

For a non-symmetric rotating harmonic trap with $\omega_x^2<\omega_y^2$, the mean-field LLL picture from Eqs.~(\ref{para}) and (\ref{RxRy}) predicts that $R_y^2$  vanishes slowly ($\propto \sqrt\epsilon$) as $ \omega_x^2 -\Omega^2 = \epsilon^2 \to 0$.  Ultimately, such behavior must cross over to a different dependence, and Aftalion {\it et al.}~\cite{Afta09} suggest the formation of a quasi one-dimensional regime with no visible vortices.  This idea arises from the LLL projector in Eqs.~(\ref{anisoPi0}) and (\ref{chi}), where the factor $\chi(x,y)$ implies a Gaussian dependence on the tightly confined variable $y$.

If the trial function is a narrow Gaussian $\psi_v(y) \propto \exp(-y^2/2\sigma^2)$ with mean width $\sigma$, the projected function $\psi_{v0}$ has minimum energy $E'$ when $\sigma \to 0$, yielding a projected function 
\begin{equation}
\label{psiv} \psi_{v0}(y) \propto \exp(-y^2/2y_{\rm min}^2),
\end{equation}
with $y_{\rm min}^2 \approx \Omega/\alpha \beta_-$~\cite{Afta09}.  Hence it is natural to assume a crossover from the mean-field parabolic state to a quasi one-dimensional state when $R_y^2 \approx Cy_{\rm min}^2$, where the choice of the numerical factor $C$  is somewhat arbitrary.

For rapid rotation with $\epsilon\to 0$, the characteristic squared dimension $y_{\rm min}^2$ approaches a constant value 
\begin{equation}
\label{ymin}
y_{\rm min}^2 = \frac{\Omega}{\alpha\beta_-} \approx \frac{1}{\sqrt2\sqrt{2-\nu^2}}.
\end{equation}
 Assuming that  the mean-field parabolic density profile in Eq.~(\ref{para})   can indeed occur in the lowest Landau level, the crossover from the parabolic parameter $R_y^2$ to the Gaussian $y_{\rm min}^2$ occurs at  a corresponding rotation rate
\begin{equation}
\label{epscross}
\epsilon_{\rm cross}^2\approx \frac{\nu^6 C^4}{1024}\,\frac{(2-\nu^2)}{(1-\nu^2)^{3}} \left(\frac{Z}{Na_s}\right)^2.
\end{equation}
With $C=10$, my previous  numerical parameters ($\nu = 0.5$, $N= 5\times 10^3$,  and $Z/a_s= 200$) imply $\epsilon_{\rm cross}^2 \approx 1.0\times  10^{-3}$, which might well be attainable.   Recent numerical studies of the formation of a single  vortex row suggests that the condensate's order parameter  remains Gaussian in this situation, so that the constant $C$ may well be as large as $20$~\cite{Afta109}.  In this  case the crossover parameter becomes  $\epsilon^2_{\rm cross} \approx 1.6\times 10^{-2}$, which should be readily observable.

Figure 2 of Aftalion {\it et al.}~\cite{Afta09} displays an effectively one-dimensional condensate with no visible vortices, but they assume a much smaller value $Na_s/a_z\approx 3$, where $a_z$ is the extension of the wave function in the $z$ direction~\cite{note2}.  Such a weak coupling constant would require a smaller particle number $N$, since $a_s$ is $\sim$ a few nm and a typical quasi one-dimensional condensate has a transverse dimension  $a_z\lesssim1 \ \mu$m.  

\section{Conclusions and outlook}

 A rapidly rotating symmetric two-dimensional condensate expands radially  and has a large squared radius $R_{LLL}^2$ proportional to $1/\epsilon$, where $\epsilon^2 = 1-\Omega^2$.   Visible quantized vortices provide the total angular momentum $L_z$.  In oscillator units, each vortex occupies an area $\approx \pi$, so that the number of vortices is $N_v\approx R_{LLL}^2$.   The criterion for a quantum phase transition to a correlated state is usually given as $N/N_v\approx f_c$.  Taking $f_c\sim 10$, I find that the critical rotation rate for this quantum phase transition is $\epsilon_Q^2 \approx 2\times 10^{-3}$, which is experimentally challenging but perhaps feasible.
 
The LLL wave function  for a symmetric trap ensures that $\langle L_z\rangle = \langle r^2\rangle -1$, and the parabolic density profile from Eq.~(\ref{dens}) yields the approximate result  $\langle L_z\rangle  \approx  \frac{1}{3}R_{LLL}^2$ for large $R_{LLL}^2$.  Hence the criterion for the quantum phase transition for a symmetric  trap  is equivalent to $N/L_z \approx 3 f_c$, which  is more general  than $N/N_v$ in cases where the angular momentum arises from other mechanisms (such as in a non-symmetric trap---see below).

A  non-symmetric trap has $\omega_y^2 -\omega_x^2 = 2\nu^2>0$, and it is now natural to define the small parameter as $\epsilon^2 =\omega_x^2 -\Omega^2$.  For rapid rotation, the condensate expands along the weak $x$ direction  and contracts along the tight  $y$ direction.  
In one scenario for a quantum phase transition, the resulting asymmetry induces a  large moment of inertia and a correspondingly  large  angular momentum $L_z$ proportional to $\epsilon^{-3/2}$.  Note that this dependence is more singular than that for a symmetric trap, where $L_z\propto \epsilon^{-1}$.  As a result, the relevant ratio $N/L_z$  for a non-symmetric trap decreases more rapidly than for a symmetric trap, favoring the quantum phase transition to a correlated state.

A second  scenario for the quantum phase transition involves a fragmented state~\cite{Dali09}. In contrast to the angular-mementum criterion, this picture implies that the quantum phase transition is more difficult to achieve for a non-symmetric trap.   

Instead of a quantum phase transition, a third  scenario predicts a smooth crossover to a quasi one-dimensional condensate with no visible vortices~\cite{Afta09}.   Previous numerical studies of symmetric traps~\cite{Coop01,Sino02,Regn03,Regn04} relied on geometries like tori or spheres that have no boundaries.  More general numerical studies that include the trap anisotropy would be highly desirable.

One intriguing experimental possibility is  to use  synthetic
 gauge potentials~\cite{Gunt09,Lin09,Spie09,Murr09}. Such  light-induced  effective vector potentials  can  mimic a rotating system while remaining in the laboratory frame.  It would be very interesting to have such experiments on a  non-symmetric trap, which could help decide among these various scenarios.


\begin{acknowledgments}
The Institut Henri Poincar{\'e}-Centre Emile Borel, Paris  (IHP) sponsored a 2007 workshop that facilitated very helpful discussions with  A.\ Aftalion, X.\ Blanc, N.\ Cooper, and N.\ Lerner.   I thank the IHP  for their hospitality and support.  I also am grateful to A.\ Aftalion,  N.~Cooper, and J.\  Dalibard for recent valuable comments and suggestions.
\end{acknowledgments}

\appendix
\section{Canonical transformation for rotating non-symmetric trap}

The Hamiltonian in Eq.~(\ref{H1}) is quadratic in the coordinates and momenta, so that a linear canonical transformation can diagonalize it, but the details become quite intricate.  Introduce a four-component vector $\bm v^T$ with elements $x,y,p_x,p_y$, where $T$ denotes a transpose of a vector.  The one-body Hamiltonian in (\ref{H1}) then has the form 
$H_0 = \frac{1}{2} \bm v^T {\cal H} \bm v$.  In the present case, the symmetric $4\times 4$ matrix ${\cal H}$ is 
\begin{equation}
 \label{Hn}
{\cal H} =
 \left( 
\begin{matrix}
\omega_x^2 &  0&0&-\Omega \\[.1cm]
0 & \omega_y^2&  \Omega&0\\[.1cm]
0&\Omega & 1&0\\[.1cm]
-\Omega&0&0&1
\end{matrix}
\right).
\end{equation}
 
 The Hamiltonian equations  of motion are equivalent to
 \begin{equation}
 \label{dynamics}
\dot{\bm v} =\sigma\frac{ \partial H}{\partial \bm v}= \sigma {\cal H} \bm v,
\end{equation}
where the  $4\times 4$  matrix 
 \begin{equation}
\sigma =\begin{pmatrix}
0&0&1&0\\0&0&0&1\\-1&0&0&0\\
0&-1&0&0
\end{pmatrix}
\end{equation}
 enforces the symplectic structure of Hamilton's equations.
I seek normal modes with time dependence $\propto e^{-i\omega t}$.  In this case, the dynamical equations (\ref{dynamics})  become a linear matrix  equation 
\begin{equation}
\label{norm}
{\cal H} \bm v = \omega i \sigma \bm v,
\end{equation}
where $i\sigma$ is a Hermitian matrix with $(i\sigma)^\dagger = i\sigma$.

The eigenvectors $\bm u_j$ obey the corresponding eigenvalue equation 
\begin{equation}
\label{evector}
{\cal H} \bm u_j = \omega_j i\sigma \bm u_j
\end{equation}
with two independent real roots $\omega_\pm$ given in Eq.~(\ref{modes}).  The resulting eigenvectors are orthogonal with respect to the matrix $i\sigma$ that serves as a metric  
\begin{equation}\label{ortho}
\bm u_j^\dagger i\sigma \bm u_k = \delta_{jk}. 
\end{equation} Straightforward but lengthy calculations yield the normalized eigenvectors 
\begin{equation}
\label{eigenvectors}
\bm u_+ = \frac{\lambda_+}{\sqrt{2\omega_+}} 
\left(\begin{matrix} 1\\[.1cm]
i\beta_+\\[.1cm]
-i(\omega_+ + \Omega \beta_+)\\[.1cm]
(\Omega+\omega_+\beta_+)
\end{matrix}\right)\quad\hbox{and}\quad 
\bm u_- = \frac{\lambda_-}{\sqrt{2\omega_-}} 
\left(\begin{matrix} i\beta_-\\[.1cm]
1\\[.1cm]
(\omega_-  \beta_- - \Omega)\\[.1cm]
i(\Omega\beta_- - \omega_-)
\end{matrix}\right),
\end{equation}
where $\lambda_\pm^{-2} = 1+\beta_+\beta_-\omega_\mp/\omega_\pm$.

Use these eigenvectors to construct the $4\times 4$ matrix  
\begin{equation}
\label{modaln}
{\cal N} =
 \left( 
\begin{matrix}
\bm u_+& \bm u_- & \bm u_+^* & \bm u_-^* \\
\downarrow & \downarrow &\downarrow &\downarrow
\end{matrix}
\right).
\end{equation}
The orthogonality relations (\ref{ortho}) are equivalent to the single matrix equation 
\begin{equation}
\label{orthogn}
{\cal N}^\dagger i\sigma {\cal N} = \tau,
\end{equation}
where $\tau$ is  a $4\times 4$-dimensional generalization of the familiar Pauli matrix $\sigma_z$ with diagonal elements ($1,1,-1,-1$). In addition, the matrix $\cal N$ 
satisfies the relation  [compare Eq.~(\ref{evector})] 
 \begin{equation}
 {\cal H N }=i\sigma  {\cal N }\tau{\cal O},
\end{equation}
where $\cal O$ is a diagonal matrix with  positive elements $\omega_+,\omega_-,\omega_+,\omega_-$.  The normal coordinates $A_+,A_-$ can  be defined through the linear transformation 
\begin{equation}
\label{evector1}
\bm v = {\cal N}\bm A, 
\end{equation}
where  the $4$-dimensional vector $\bm A$ has the explicit form (here given as the transpose)
 \begin{equation}
\bm A^T  =
 \left(
A_+,
A_-,
 A_+^*,
 A_-^*,
\right).
\end{equation}
This transformation also diagonalizes the original Hamiltonian
\begin{equation}
H =\frac{1}{2} \bm A^\dagger {\cal O}\bm A = \frac{1}{2} \omega_+\left(A_+^\dagger A_+ + A_+ A_+^\dagger\right) +\frac{1}{2} \omega_-\left(A_-^\dagger A_-+ A_-A_-^\dagger\right).
\end{equation}

Equations (\ref{orthogn})  and (\ref{evector1}) yield the explicit form of the normal coordinates 
\begin{equation}
\label{normaln}
\bm A = \tau {\cal N}^\dagger i\sigma \bm v,
\end{equation}
 in terms of the  eigenvectors $\bm u_\pm$ and the original canonical coordinates $x,y,p_x,p_y$.   A detailed analysis yields the rather complicated expressions 
 \begin{eqnarray}
\label{ap}
A_+ = \frac{\lambda_+}{\sqrt{2\omega_+}}\left[ (\omega_+ + \Omega\beta_+)x +\beta_+p_y - i (\Omega+\beta_+\omega_+)y + i p_x\right], \\[.2cm]
\label{am}
A_- = \frac{\lambda_-}{\sqrt{2\omega_-}}\left[ (\omega_- -\Omega\beta_-)y +\beta_-p_x +  i (\Omega- \beta_-\omega_-)x + i p_y\right],
\end{eqnarray}
that  generalize the results (\ref{operators}) for a rotating symmetric trap to the case of  a rotating non-symmetric trap.  In the symmetric limit ($\nu \to 0$), $A_+\to a_+$, but $A_-  \to -i a_-=(a_y-ia_x)/\sqrt 2$, which is merely an additional canonical transformation.


\begin{thebibliography}{99}
\bibitem{note1a}  A more careful study  includes the energy associated with the formation of the vortex cores within each unit cell.   This effect renormalizes the two-dimensional coupling constant by a numerical factor.  Equivalently,  the  $s$-wave scattering length is rescaled $a_s \to b a_s$, where $b \approx 1.1596$~\cite{Afta05}.  Since I here consider only qualitative estimates,  such renormalization is unimportant for my purposes.
\bibitem{note2} Note that Ref.~\cite{Afta09} omits a numerical factor  $\sqrt{8\pi}$  in their definition of the dimensionless coupling constant $G = Na_s/a_z$~\cite{note1}.  For my choice of a condensate with uniform  thickness $Z$, the dimensionless coupling constant is generally taken as $4\pi N a_s/Z$~\cite{Rose02}.
\end{thebibliography}
\end{document}